\documentstyle[12pt]{article}

\begin{document}

\title{Dissipative cosmological solutions}

\author{Luis P. Chimento and Alejandro S. Jakubi      \\
{\it Departamento de F\'{\i}sica,  }\\
{\it  Facultad de Ciencias Exactas y Naturales, }\\
{\it Universidad de Buenos Aires }\\
{\it  Ciudad  Universitaria,  Pabell\'{o}n  I, }\\
{\it 1428 Buenos Aires, Argentina.}}

\maketitle

\begin{abstract}

The exact general solution to the Einstein equations in a homogeneous Universe
with a full causal viscous fluid source for the bulk viscosity index $m=1/2$
is found. We have investigated the asymptotic stability of Friedmann and de
Sitter solutions, the former is stable for $m\ge 1/2$ and the latter for $m\le
1/2$. The comparison with results of the truncated theory is made. For
$m=1/2$, it was found that families of solutions with extrema no longer remain
in the full case, and they are replaced by asymptotically Minkowski
evolutions. These solutions are monotonic.

\end{abstract}

\vskip 3cm


\newpage

\section{Introduction}

It is believed that quantum effects played a fundamental role in the early
Universe. For instance, vacuum polarisation and particle production arise from
a quantum description of matter. However it is known that both of them can be
modeled in terms of a classical bulk viscosity \cite{Hu82}.
Other processes capable of producing important dissipative stresses
include interactions between matter and radiation
\cite{Wei71}, quarks and gluons \cite{Tho}, different components of dark
matter \cite{Pav93}, and those mediated by massive particles. It happens also
due to the decay of massive superstrings modes \cite{Myu}, gravitational
string production \cite{Tur88} \cite{Bar88} and phase transitions.

Cosmological models with a viscous fluid have been studied by several authors.
Some of the interesting subjects addressed by them were the effects of viscous
stresses on the avoidance of the initial singularity \cite{Mur}, the
dissipation of a primordial anisotropy \cite{Mis}, the production of entropy
\cite{Wei71}, inflation and deflation \cite{Bar86}. The evolution of a
homogeneous isotropic spatially-flat universe filled with a causal viscous
fluid, whose bulk viscosity coefficient is related to the energy density by
the power law $\zeta \sim \rho^m$, and a truncated version of the transport
equation for the viscous pressure, has been investigated thoroughly \cite{Pav}
\cite{Zak93a} \cite{Chi93}. This model was based in the relativistic
second-order theory of non-equilibrium thermodynamics developed by Israel and
Steward \cite{Isr76}. Their formulation employs fiducial local equilibrium
magnitudes like thermodynamical pressure and temperature, related by an
equilibrium Gibbs equation, but generalizes the expression of the entropy flux
by the inclusion of quadratic terms in dissipative non-equilibrium magnitudes
like bulk-viscosity pressure and heat flux. An alternative formulation of this
theory, called Extended Irreversible Thermodynamics has been made by Pav\'on,
Jou and co-workers \cite{Pav82} \cite{Jou93}. Here, non-equilibrium magnitudes
like temperature and pressure are introduced instead, and a generalized Gibbs
equation including the dissipative magnitudes is employed. The equivalence
between both formalisms up to second-order departure from equilibrium has been
demonstrated \cite{Gar94}.

Recently, it was shown that the truncated version of the transport equation
leads to pathological behaviour in the late universe, as the thermodynamical
temperature increases with the expansion, leading to an unphysical heating of
the Universe \cite{Maa95}. Then, some authors have begun to employ the full
version of the transport equation for the bulk viscous pressure, and made
comparisons with the results obtained with the truncated one \cite{His91}
\cite{Zak93b} \cite{Rom} \cite{Maa95} \cite{Men96} \cite{Zim96} \cite{Col96}
\cite{Men96b} \cite{Maa97}. One important issue that
arises in this context is the choice of the equations of state. For a
dissipative Boltzmann gas it was shown that inflationary solutions appear not
to arise, and it was conjectured that the inflationary solutions of the
truncated causal theory are spurious and an artificial consequence of
truncation \cite{His91}. In \cite{Chi96} we have adopted also the full version
of the transport equation but we have taken the simple forms of the equations
of state used in \cite{Rom} and \cite{Maa95}, and we have classified the
possible behaviors near a singularity. In the present paper we complete the
analysis of this cosmological model. We present the set of equations that
describe the model in section \ref{s2}. In section \ref{s3} we solve them when
$m=1/2$ and give a detailed analysis of their solutions, including an analysis
of the energy conditions \cite{Haw}. In section \ref{s4} we study of stability
of those asymptotic solutions that appear for $m\ne 1/2$ in the truncated
causal and noncausal theories by means of the Lyapunov method.  The
conclusions are stated in section \ref{s6}.

\section{The Model} \label{s2}

In the case of the homogeneous, isotropic, spatially flat Robertson-Walker
metric 

\begin{equation} \label{1}
ds^{2} = dt^{2} - a^{2}(t) ( dx^{2}_{1} + dx^{2}_{2} + dx^{2}_{3} ) 
\end{equation}

\noindent only the bulk viscosity needs to be considered. Thus we replace in
the Einstein equations the equilibrium pressure $p$ by an effective pressure 
\cite{Wei71} 

\begin{equation} \label{2}
H^{2} = {\frac{1}{3}} \rho
\end{equation}

\begin{equation} \label{3}
\dot H+ 3H^{2} = {\frac{1}{2}} (\rho - p -\sigma )
\end{equation}

\noindent where $H$= $\dot a/a$, $^{\cdot }=d/dt$, $\rho $ is the energy
density, $\sigma $ is the viscous pressure, and we use units $c=8\pi G=1$.
As equation of state we take

\begin{equation} \label{4}
p = ( \gamma -1 ) \rho  
\end{equation}

\noindent  with a constant adiabatic index $0\le\gamma \le 2$, and for $\sigma
$ we assume the transport equation,

\begin{equation} \label{4.1}
\sigma+\tau\dot\sigma=-3\zeta H-\frac{1}{2}\tau\sigma
\left(3H+\frac{\dot\tau}{\tau}-\frac{\dot\zeta}{\zeta}-\frac{\dot T}{T}\right)
\end{equation}

\noindent which arises from the causal irreversible thermodynamics theory, as
the simplest way (linear in $\sigma$) to guarantee positive entropy production
\cite{Maa95} \cite{Zak93b}. Here $\zeta$ is the bulk viscous coefficient, and
following \cite{Bel75}

\begin{equation} \label{4.2}
\zeta = \alpha \rho ^{m}
\end{equation}

\noindent where $\alpha>0 $ and $m$ are constants. Also, $\tau$ is the bulk
relaxation time and  we choose the law 

\begin{equation} \label{5}
\tau = \frac{\zeta}{\rho}
\end{equation}

\noindent which ensures causal propagation of viscous signals \cite {Bel79}.
Finally, $T$ is the equilibrium temperature and following \cite{Rom} we choose

\begin{equation} \label{6}
T=\kappa\rho^r
\end{equation}

\noindent
where $\kappa>0$ and $r\ge 0$ are constants, which is the simplest
way to guarantee a positive heat capacity. In the context of
standard irreversible thermodynamics, $p$, $\rho$, $T$ and the number density
$n$ are equilibrium magnitudes, which are related by equations of state of the
form $\rho=\rho(T,n)$ and $p=p(T,n)$. Further the
thermodynamic relation holds \cite{Pav93}

\begin{equation} \label{6.1}
\left(\frac{\partial\rho}{\partial n}\right)_{T}=\frac{\rho+p}{n}-
\frac{T}{n}\left(\frac{\partial p}{\partial T}\right)_{n},
\end{equation}

\noindent
which follows from the requirement that the entropy is a state function. In our
model, this relation imposes the constrain

\begin{equation} \label{7}
r=\frac{\gamma-1}{\gamma}
\end{equation}

\noindent
so that $0\le r\le 1/2$ for $1\le\gamma\le 2$.
Using (\ref{2})-(\ref{6}), we obtain

$$
\ddot H-(1+r)\frac{\dot H^2}{H}+
\frac{3}{2}\left[1+\left(1-r\right)\gamma\right] H\dot H+
\frac{3^{1-m}}{\alpha}H^2 \left|H\right|^{-2m}\dot H+
$$

\begin{equation} \label{8}
\frac{9}{4}\left(\gamma-2\right) H^3+
\frac{3^{2-m}}{2\alpha}\gamma H^4 \left|H\right|^{-2m}=0
\end{equation}

\noindent   
for $m\neq 1/2$, and

\begin{equation} \label{9}
H\ddot H-\left(1+r\right)\dot H^2+A H^2\dot H-B H^4=0
\end{equation}

\noindent for $m=1/2$, where
$A\equiv 3/2\left[1+\left(1-r\right)\gamma+2/\gamma_0\right]$,
$B\equiv 9/4\left(2-\gamma-2\gamma/\gamma_0\right)$, and
$\gamma_0=\sqrt{3}\alpha\, {\rm sgn}H$. It is easy to check that solutions of
(\ref{9}) never change sign, and we will consider in the following only
expanding evolutions, that is $H>0$. Thus, taking into account (\ref{7}), we
note that $A>0$.

\section{The case $m=1/2$}\label{s3}

For the integration of (\ref{9}), provided that $r\neq 0$,
it is convenient to make the change of variable $H=y^{-1/r}$. It turns this
equation into

\begin{equation} \label{10}
\ddot y+A y^{-\frac{1}{r}}\dot y+Bry^{1-\frac{2}{r}}=0
\end{equation}

\noindent
which is a particular instance of the class of equations

\begin{equation} \label{101}
\ddot y+ \alpha_1 f(y)\dot y+\alpha_2 f(y)\int dy f(y)+\alpha_3 f(y)=0
\end{equation}

\noindent that linearises through the transformation \cite{Chi93} \cite{Chi96}
\cite{Chi96b} \cite{Chi96c}

\begin{equation} \label{11}
z=\int dy f(y),\qquad
\eta=\int dt f(y)
\end{equation}

\noindent
becoming

\begin{equation} \label{111}
\frac{d^2z}{d\eta^2}+\alpha_1\frac{dz}{d\eta}+\alpha_2 z+\alpha_3=0
\end{equation}

\noindent Thus, for $\alpha_1=1$, $\alpha_2=(r-1)B/A^2$ and $\alpha_3=0$ we
obtain the general solution of (\ref{9}) in parametric form

$$
H(\eta)=\left[Rz(\eta)\right]^{\frac{1}{1-r}}
$$

\begin{equation} \label{12}
\Delta t(\eta)=\frac{1}{A}\int d\eta H(\eta)^{-1}
\end{equation}

\noindent
where $\Delta t=t-t_0$, and $t_0$ is an arbitrary constant and $R$ is a
constant. For the case $r=0$, that is isothermal
matter, it is easy to verify that (\ref{12}) is also the general solution. The
constrain (\ref{7}) implies that

\begin{equation} \label{13}
z(\eta)=C_+\exp\left(\lambda_+\eta\right)+
C_-\exp\left(\lambda_-\eta\right)
\end{equation}

\noindent
where $\lambda_\pm=(1/2)[-1\pm\left(1-4\alpha_2\right)^{1/2}]$ are real,
and $C_\pm$ are arbitrary integration constants. When any one of  them vanishes,
we obtain the one-parameter families of solutions, which have explicit form:

\begin{equation}
\label{10a}
H_{\pm }(t)=\nu _{\pm }/\Delta t,\qquad B \neq 0
\end{equation}
\begin{equation}
\label{10b}
H_{+}=\nu _0/\Delta t,\qquad H_{-}=H_0,\qquad B =0
\end{equation}
\begin{equation} \label{10c}
\nu _{\pm }={\frac{1}{2B}}\left\{ -A\pm
\left[ A^2-4B\left(r-1\right)\right]^{1/2}\right\}\qquad
\nu _0=\frac{1-r}{A}
\end{equation}

\noindent and $H_0$ is an arbitrary positive constant. As $0\le\nu_+\le
(2/3)(\sqrt{2}-1)$ and $1/6\le\nu_0\le2/9$, the behavior of $a_+(t)$ is
Friedmann. On the other hand, the de Sitter behavior for $B=0$ has already
been considered in \cite{Maa95} \cite{Men96}.

\subsection{Two-parameter families of solutions}

As shown in appendix A, for some values of $\gamma$ and $\gamma_0$, $a(t)$ can
be written in closed form in terms of known functions. However, for arbitrary
values of $\gamma$ and $\gamma_0$, we need to study the solution in the
parametric form (\ref{12}). Depending on the case, the behavior of $H(t)$ for
small or large time is given by the behavior of $z(\eta)$ near a zero point or
for $\pm\infty$. In the first case, $z\sim\eta$ for $\eta\to 0$, $H\sim
1/(\Delta t)^{1/r}$, provided $r\neq 0$, or $H\sim \exp(H\Delta t)$ if $r=0$.
Thus

$$
a(t)\sim a_0\exp\left(K \Delta t^{\frac{r-1}{r}}\right),\qquad r>0
$$

\begin{equation} \label{15}
a(t)\sim a_0\exp\left[\frac{1}{K}\exp\left(K\Delta t\right)\right],
\qquad r=0
\end{equation}

\noindent where $a_0$ and $K$ are arbitrary integration constants. In the
second case, either the first or the second term in (\ref{13}) dominates, and
the leading behavior of $H(t)$ is given by (\ref{10a}) (or (\ref{10b}), if
$B=0$). Thus, we obtain the following classification of the two-parameter
families of solutions:

{\bf A.} The evolution begins at a singularity with a Friedmann leading
behavior as (\ref{10a}+) for $B\neq 0$ or (\ref{10b}+) for $B=0$, and so
 there are particle horizons. Then the scale
factor expands with an asymptotically Minkowski behavior for large time like
(\ref{15}).

{\bf B.} The evolution is asymptotically Minkowski in the far past with
behavior (\ref{15}), then it expands and its behavior in the future is either:

1. asymptotically Friedmann, as (\ref{10a}--) for $B<0$;

2. asymptotically de Sitter, as (\ref{10b}--) for $B=0$;

3. divergent at finite time, with leading behavior (\ref{10a}--) for $B>0$.

{\bf C.} The evolution begins at a singularity with a Friedmann leading
behavior as (\ref{10a}+) for $B\neq 0$ or (\ref{10b}+) for $B=0$; and so there
are particle horizons. Then the scale factor expands, and its behavior in the
future is like B.

\subsection{Energy conditions}

Let us consider whether this viscous fluid obeys the energy conditions when
$m=1/2$ \cite{Haw}. Using (\ref{2})(\ref{3}), the dominant energy condition
(DEC) $\rho \ge |p+\sigma|$ implies $-3H^2\le\dot H \le 0$, while the strong
energy condition (SEC) $\rho +3p+3\sigma \ge 0$ implies $\dot H+H^{2}\le 0$.
We find that DEC is violated part of the time in the following families:  B1,
in the far past as well as for large times (if $\nu _{-}<1/3$); near the
singularity in C1, C2 and C3; C3, near the "explosion". DEC is violated always
in families A, B2 and B3.
SEC is satisfied always in families A and C1 (if $\nu _{-}\le 1$). It is
violated part of the time in the families: B1, in the far past (if $\nu
_{-}\le 1$); for large times in C1 (if $\nu _{-}>1$), C2 and C3. SEC is
violated always in families B1 (if $\nu _{-}>1$), B2 and B3. Thus, we find
that all these two-parameter solutions violate the energy conditions sometime.

\section{The case $m\neq 1/2$}\label{s4}

We will make use of the method of the Lyapunov function \cite{Kra} to
investigate the asymptotically stability of the de Sitter and asymptotically
Friedmann solutions that occur in the noncausal and truncated causal models.

\subsection{ Stability of the de Sitter solution}

Equation (\ref{8}) admits a de Sitter solution for $\gamma<2$ \cite{Maa95}

\begin{equation} \label{16}
H_0=\left[3^m\alpha\left(\frac{2-\gamma}{2\gamma}\right)\right]
^{\frac{1}{1-2m}}
\end{equation}

\noindent To study its stability, if $r>0$, we make first the change of
variable $H=y^{-1/r}$ in (\ref{8})

$$
\ddot y+\frac{3}{2}\left[1+\left(1-r\right)\gamma\right]y^{-\frac{1}{r}}\dot y
+\frac{3^{1-m}}{\alpha}y^{\frac{2}{r}(m-1)}\dot y+
$$

\begin{equation} \label{16.1}
\frac{9}{4}\left(2-\gamma\right)ry^{1-\frac{2}{r}}-
\frac{3^{2-m}\gamma}{2\alpha}ry^{1-\frac{3}{r}+\frac{2m}{r}}=0
\end{equation}

\noindent
and then we rewrite it as

\begin{equation} \label{17}
\frac{d}{dt}\left[\frac{1}{2}\dot y^2+V(y)\right]=
-\left\{\frac{3}{2}\left[1+\left(1-r\right)\gamma\right]y^{-\frac{1}{r}}+
\frac{3^{1-m}}{\alpha}y^{\frac{2(m-1)}{r}}\right\}\dot y^2
\end{equation}

\noindent
where

\begin{equation} \label{18}
V(y)=\frac{9\left(2-\gamma\right)r^2}{8\left(r-1\right)}y^{\frac{2(r-1)}{r}}-
\frac{3^{2-m}\gamma r^2}{2\alpha(2r-3+2m)}y^{\frac{2r-3+2m}{r}}
\end{equation}

\noindent
for $2r-3+2m\neq 0$, and

\begin{equation} \label{19}
V(y)=\frac{9\left(2-\gamma\right)r^2}{8\left(r-1\right)}y^{\frac{2(r-1)}{r}}-
\frac{3^{\frac{1}{2}+r}\gamma r}{2\alpha}\ln \frac{y}{y_0}
\end{equation}

\noindent for $2r-3+2m= 0$. We see that this potential has a unique minimum
for $y>0$ at $y_0=H_0^{-r}$ provided $m<1/2$, while it has a maximum for
$m>1/2$. Also, taking into account
(\ref{7}), the right hand side of (\ref{17}) is negative definite.

On the other hand, if $r=0$, we make the change of variable $dt=dx/H$ in
(\ref{8}), and we obtain

\begin{equation} \label{19.1}
\frac{d}{dx}\left[\frac{1}{2}H'^2+V(H)\right]=
-H'^2\left[3+\frac{3^{1-m}}{\alpha}H^{1-2m}\right]
\end{equation}

\noindent
where $'\equiv d/dx$, and

\begin{equation} \label{19.2}
V(H)=-\frac{9}{8}H^2+
\frac{3^{2-m}}{2\alpha\left(3-2m\right)}H^{3-2m}, \qquad m\neq \frac{3}{2}
\end{equation}

\begin{equation} \label{19.3}
V(H)=-\frac{9}{8}H^2+
\frac{\sqrt{3}}{2\alpha}\ln\frac{H}{H_0},\qquad m=\frac{3}{2}
\end{equation}

\noindent
This potential has a unique minimum
for $H>0$ at $H_0$ provided $m<1/2$, while it has a maximum for
$m>1/2$. Also the right hand side of (\ref{19.1}) is negative definite.

Thus we find
that an exponential inflationary regime is asymptotically stable for
$t\to\infty$ for any initial condition $H>0$ provided that $m<1/2$, but this
regime becomes unstable for $m>1/2$. This result improves over previous
studies based in small perturbations about solution (\ref{16}) \cite{Maa95}
\cite{Men96}.

\subsection{Stability of the asymptotically Friedmann solution}

For $m>1/2$ it is easy to check that (\ref{8}) admits a solution whose leading
term is $2/(3\gamma t)$. To study its stability, if $r>0$, we make  the change
of variables

\begin{equation} \label{20}
H = \frac{\left[u(z)\right]^{-\frac{1}{r}}}{t},\qquad t^{2m-1} = z
\end{equation}

\noindent
in (\ref{8}) which takes the form

\begin{equation}
\label{21}
{\frac{d}{dz}}\left[{\frac{1}{2}}{u^{\prime}}^{2}+U(u,z)\right]=D(u,u',z)
\end{equation}

\noindent
where $'\equiv d/dz$. We consider that $u$ lies in a neighbourhood of
$u_0\equiv\left(3\gamma/2\right)^{-1/r}$. Thus, when $z\to\infty$

\begin{equation} \label{22}
U(u,z)=\frac{3^{1-m}r^2}{2\alpha(2m-1)^2}\left[\frac{u^{\frac{2(r+m-1)}{r}}}
{r+m-1} -
\frac{3\gamma u^{\frac{2r+2m-3}{r}}}{2r+2m-3}
\right]\frac{1}{z}+O\left(\frac{1}{z^2}\right)
\end{equation}

\noindent
for $2r+2m\neq 2,3$,

\begin{equation} \label{22.1}
U(u,z)=\frac{3^{r}r}{\alpha(1-2r)^2}\left(\ln\frac{u}{u_0}+
\frac{3\gamma r}{2}u^{-\frac{1}{r}}\right)\frac{1}{z}+
O\left(\frac{1}{z^2}\right),\qquad r+m= 1
\end{equation}

\begin{equation} \label{22.2}
U(u,z)=\frac{3^{r-\frac{1}{2}}r}{4\alpha(1-r)^2}\left(ru^{\frac{1}{r}}-
\frac{3\gamma}{2}\ln\frac{u}{u_0}\right)\frac{1}{z}+
O\left(\frac{1}{z^2}\right),\qquad 2r+2m=3
\end{equation}

\noindent
and

\begin{equation} \label{23}
D\left(u,u',z\right)=-\frac{3^{1-m}}{\alpha(2m-1)}u^{\frac{2(m-1)}{r}}u'^2+
O\left(\frac{1}{z}\right)
\end{equation}

On the other hand, if $r=0$, we make the change of variables

\begin{equation} \label{23.1}
H = \frac{\exp\left[u(z)\right]}{t},\qquad z=t^{2m-1}
\end{equation}

\noindent
in (\ref{8}) which takes again the form (\ref{21}), but now
we consider that $u$ lies in a neighbourhood of
$u_0\equiv\ln(2/3)$, and when $z\to\infty$

\begin{equation} \label{23.3}
U(u,z)=\frac{3^{1-m}}{2\alpha\left(2m-1\right)^2}
\left[\frac{3 e^{(3-2m)u}}{3-2m}-\frac{e^{(2-2m)u}}{1-m}\right]
\frac{1}{z}+O(\frac{1}{z^2})
\end{equation}

\noindent
for $2m\neq2,3$,

\begin{equation} \label{23.4}
U(u,z)=\frac{1}{\alpha}\left(\frac{3}{2}e^u-u\right)
\frac{1}{z}+O(\frac{1}{z^2}),\qquad m=1
\end{equation}

\begin{equation} \label{23.5}
U(u,z)=\frac{1}{4\sqrt{3}\alpha}\left(\frac{3}{2}u+e^u\right)
\frac{1}{z}+O(\frac{1}{z^2}),\qquad 2m=3
\end{equation}

\noindent
and

\begin{equation} \label{23.6}
D(u,u',z)=-\frac{3^{1-m}}{\alpha(2m-1)}e^{(2-2m)u}u'^2+
O\left(\frac{1}{z^2}\right)
\end{equation}

As $U(u,z)$ has a unique minimum at $u_0$ for any $m\neq 1/2$,
and $D(u,u',z)$ is negative definite for $m>1/2$, we find that solutions
with leading Friedmann behaviour $a\sim t^{2/(3\gamma)}$ when $t\to\infty$ are
asymptotically stable for $m>1/2$ , but become unstable for $m<1/2$.

\section{Conclusions}\label{s6}

We have improved our previous analysis of a cosmological model with a causal
viscous fluid by considering the full transport equation for the bulk
viscosity instead of the truncated one \cite{Chi93}. We have kept the equation
of state that relates the equilibrium pressure and energy density, and we have
added a power-law relationship between the equilibrium temperature and energy
density. 

When $m=1/2$, the splitting of the large time asymptotic behavior of solutions
in terms of sgn$(-B)$ follows closely that of the truncated model in terms of
sgn $(\gamma-\gamma_0$ ), which in its turn resembles the classification for
the noncausal solutions. However, families of solutions with extrema no longer
remain in the full case, and we find instead asymptotically Minkowski
evolutions. Moreover, all solutions are monotonic.

The singular behavior for $m=1/2$ is also very similar to that of the
truncated model because, excluding (\ref{10a}--), singular solutions have
particle horizons. The only difference is that $\nu_+$ in (\ref{10c}+) is
smaller than the corresponding magnitude in \cite{Chi93}. Similar result has
already been verified for $m>1/2$ \cite{Chi96}.

As in the truncated model, we demonstrate that there is no value of $m$ for
which there is a stable expanding de Sitter period in the far past. This
further support the conclusion of that causality avoids the deflationary
behavior proposed by Barrow \cite{Bar88}.

If $m<1/2$, a stable exponential inflationary phase occurs in the far future
for $ 1\le\gamma<2$; the condition $B=0$ is required if $m=1/2$, and such a
behavior is unstable for $m>1/2$. This is the same structural behavior already
found in the causal truncated and noncausal models. It has been noted however
that the inflation rate in the full causal theory is different, lower for
$m<1/2$ and greater for $m>1/2$. Also, there is no stiff-fluid viscous
exponential inflation, unlike in the truncated or non-causal theories
\cite{Maa95}.

It has been conjectured that viscosity-driven exponential inflationary
solutions are spurious and an artificial consequence of use of causal
truncated  or noncausal theories \cite{His91}. This suggestion arises from
study of a cosmological model containing a dissipative Boltzmann gas. Our
results show however that this kind of conclusions does not generalize to
other models where the viscous fluid is described by a different equation of
state. Moreover, they rather suggest that some features of the solutions, like
their asymptotic behavior depend more strongly on the equation of state than
on the thermodynamical theory employed.

If $m>1/2$, we find that the perfect fluid behavior $a\sim t^{2/(3\gamma)}$
for $t\rightarrow\infty$ is asymptotically stable.  This arises because the
viscous pressure decays faster than the thermodynamical pressure. However, if
$m=1/2$ and $B<0$, both pressures decay asymptotically as $t^{-2}$ and the
exponent becomes $\nu_-$. The perfect fluid behavior becomes unstable if
$m<1/2$. Here also, no structural change occurs with respect to the truncated
causal or noncausal models.

\section{Appendix A}

The equation (\ref{9}) has a two-dimensional Lie group of point symmetries,
whose generators are $\partial/\partial t$ and $t\partial/\partial t-
H\partial/\partial H$. We look for a simple transformation to a new
dependent variable $v$ that is linear in $\dot v$, preserves this group of
symmetries and keeps the order of the differential equation. This technique
can be extended to some polynomial autonomous differential equations arising
when we consider quantum effects due to vacuum polarisation terms
\cite{Chi88}. Thus, we are led to the transformation

\begin{equation} \label{A0}
\dot H=f H^2+g H \frac{\dot v}{v}
\end{equation}

\noindent where $f$ and $g$ are two constants determined by the requirement
that the differential equation in $v(t)$ is linear. Inserting (\ref{A0}) in
(\ref{9}) we obtain

$$
\left[\left(1-r\right)f^2+Af-B\right]H^4-g\left[\left(2r-1\right)f-A\right]
\frac{\dot v}{v}H^3-
$$

\begin{equation} \label{A01}
g\left[-\frac{\ddot v}{v}+\left(gr+1\right)
\frac{{\dot v}^2}{v^2}\right]H^2=0
\end{equation}

\noindent Then, provided we choose $f=A/\left(2r-1\right)$, $g=-1/r$, and we
impose the constrain

\begin{equation} \label{A1}
\frac{Br}{A^2}-\frac{1}{\left(2-\frac{1}{r}\right)^2}=0
\end{equation}

\noindent
equation (\ref{9}) turns into $\ddot v=0$. To integrate (\ref{A0}), we make
the change of variable $H=-\dot u/\left(fu\right)$, an it becomes

\begin{equation} \label{A11}
\frac{\ddot u}{u}+\frac{1}{r}\frac{\dot v}{v}=0
\end{equation}

\noindent
which has the first integral $\dot uv^{-1/r}=C$, with some arbitrary
integration constant $C$. Thus we obtain

\begin{equation} \label{A1.1}
H=\left(1-2r \right)\frac{v^{-\frac{1}{r}}}{A\int dt
v^{-\frac{1}{r}}}
\end{equation}

\noindent
and the scale factor

\begin{equation} \label{A2}
a(t)=a_0\left|\left|\Delta t\right|^{\frac{r-1}{r}}+K\right|^{\nu}
\end{equation}

\noindent where $\nu=(1-2r)/A$ and $K$ is an arbitrary integration constant.
We note that, using (\ref{7}), (\ref{A1}) becomes a curve on the parameter
plane $(\gamma,\gamma_0)$. This explicit general solution can
also be derived  from a more comprehensive mathematical framework, as shown
in \cite{Chi96b}.

For $B=0$, an implicit expression for the
solution of (\ref{9}) exists. In effect, equation (\ref{10}) admits a
first integral

\begin{equation} \label{A3}
\dot y+\frac{Ar}{r-1}y^{\frac{r-1}{r}}=E
\end{equation}

\noindent
and we obtain

\begin{equation} \label{A4}
\Delta t=\frac{1}{H_0}\left(\frac{H_0}{H}\right)^r
 {}_2F_1\left(1,\frac{r}{r-1},
\frac{2r-1}{r-1},\frac{Ar}{r-1}\left(\frac{H}{H_0}\right)^{1-r}\right)
\end{equation}

\begin{equation} \label{A5}
\Delta t=\frac{(1-r)}{AH_0}\left(\frac{a}{a_0}\right)^{\frac{A}{1-r}}
{}_2F_1\left(\frac{1}{1-r},\frac{1}{1-r},\frac{2-r}{1-r},
\left(\frac{a}{a_0}\right)^A\right)
\end{equation}

\noindent
where $H_0$, $a_0$ are arbitrary integration constants.

\newpage

{\LARGE Figure Captions}

\vskip 1.5cm

{\bf Fig. 1.} Selected solutions for $\gamma=1.1$ ($B>0$).
The curves of class A and C3 correspond to (\ref{A2}) with  $K=-1$.
The curve of class B3 corresponds to (\ref{A2}) with  $K=1$.

\end{document}